\documentstyle[twocolumn,aps,epsfig,psfig,citesort]{revtex}

\begin{document}

\def \lesssim{\leavevmode{\raisebox{-.5ex}{ $\stackrel{<}{\sim}$ } } }
\def \gtrsim{ \leavevmode{\raisebox{-.5ex}{ $\stackrel{>}{\sim}$ } } }
\def \subarr{  \begin{array}{c} \mbox{\scriptsize{$ \{ n_{\ell}\}$ }} \\ 
        \mbox{\scriptsize{ $\{m_{\ell}\}$}}             \\ 
        \mbox{\scriptsize{ $\{p_{\ell}\}$ }}  \end{array} }
\def \subun{ \leavevmode{\raisebox{-1.6ex}{$\stackrel{
	\textstyle{\ell_i,\ell_j}}
	{\mbox{\scriptsize{unconstrained}}} $}} }
\def \maxj{\begin{array}{c} \mbox{max} \\ 
        \mbox{\scriptsize{ $j$}}             \\ 
        \mbox{\scriptsize{ $ 0 \leq j \leq \jmax $ }}  \end{array} }
\def \minF{\begin{array}{c} \mbox{min} \\ 
        \mbox{\scriptsize{ $F$}}  \end{array} }
\def \minNfo{\begin{array}{c} \mbox{min} \\ 
        \mbox{\scriptsize{ $\Nfo$}}  \end{array} }
\def \nl{ \textstyle{\mbox{\scriptsize{$\{n_{\ell}\}$}}} }
\def \ml{ \mbox{\scriptsize{$\{m_{\ell}\}$}} }
\def \pl{ \mbox{\scriptsize{$\{p_{\ell}\}$}} }
\def \be{\begin{equation}}
\def \ee{\end{equation}}
\def \bea{\begin{eqnarray}}
\def \eea{\end{eqnarray}}
\def \zn{z_{\mbox{\tiny N}}}
\def \jmax{j_{\mbox{\tiny MAX}}}
\def \ei{\epsilon_i}
\def \ej{\epsilon_j}
\def \Qj{Q_j}
\def \Qk{Q_k}
\def \Qi{Q_i}
\def \Qu{Q_{\mbox{\tiny U}}}
\def \Qsurf{Q_{\mbox{\tiny SURF}}}
\def \eio{\epsilon_i^o}
\def \ejo{\epsilon_j^o}
\def \eabn{\epsilon_{\alpha\beta}^{\mbox{\tiny N}}}
\def \eab{\epsilon_{\alpha\beta}}
\def \Dab{\Delta_{\alpha\beta}}
\def \Dabn{\Delta_{\alpha\beta}^{\mbox{\tiny N}}}
\def \setDab{\{ \Delta_{\alpha\beta} \} }
\def \setDabn{\{ \Delta_{\alpha\beta}^{\mbox{\tiny N}} \} }
\def \li{\ell_i}
\def \leff{\ell_{\mbox{\tiny EFF}}}
\def \lj{\ell_j}
\def \si{s_i}
\def \sj{s_j}
\def \sm{s_m}
\def \be{\begin{equation}}
\def \ee{\end{equation}}
\def \bea{\begin{eqnarray}}
\def \eea{\end{eqnarray}}
\def \aQ{\alpha_{\mbox{\tiny Q}}}
\def \muQ{\mu_{\mbox{\tiny Q}}}
\def \lQ{\lambda_{\mbox{\tiny Q}}}
\def \lT{\lambda T}
\def \D{\partial}
\def \DF{\Delta F}
\def \DFdag{\Delta F^{\neq}}
\def \Fdag{F^{\neq}}
\def \Fbardag{\overline{F}^{\neq}}
\def \fbar{\overline{f}}
\def \dfi{\delta f_i}
\def \Nfdag{N_{\mbox{\tiny F}}^{\neq}}
\def \Nfo{N_{\mbox{\tiny F}}^{o}}
\def \Nf{N_{\mbox{\tiny F}}}
\def \nf{n_{\mbox{\tiny F}}}
\def \nfdag{n_{\mbox{\tiny F}}^{\neq}}
\def \Fbar{\overline{F}}
\def \Fstar{F^{\star}}
\def \dij{\delta_{ij}}
\def \di{\delta_{i}}
\def \dj{\delta_{j}}
\def \d{\delta}
\def \s{\sigma}
\def \Qdag{Q^{\neq}}
\def \Qddag{Q^{\neq}}
\def \Qo{Q^{o}}
\def \En{E_{\mbox{\tiny N}}}
\def \dli{\delta\ell_i}
\def \dl{\delta\ell}
\def \dlj{\delta\ell_j}
\def \dei{\delta\epsilon_i}
\def \dsi{\delta s_i}
\def \de{\delta\epsilon}
\def \dej{\delta\epsilon_j}
\def \dQi{\delta Q_i}
\def \dQj{\delta Q_j}
\def \dQ{\delta Q}
\def \ds{\delta s}
\def \setei{\left\{ \epsilon_i \right\}}
\def \eistar{\ei^{\star}}
\def \setsi{\left\{ s_i \right\}}
\def \setej{\left\{ \epsilon_j \right\}}
\def \setli{\left\{ \ell_i \right\}}
\def \setlj{\left\{ \ell_j \right\}}
\def \setQi{\left\{ Q_i \right\}}
\def \setQiQ{\left\{\Qi\left(Q\right)\right\}}
\def \setQiQstar{\left\{\Qi^{\star}\left(Q\right)\right\}}
\def \setQistarQ{\left\{ Q_i^{\star} \left(Q\right)\right\}}
\def \setQistar{\left\{ Q_i^{\star} \right\}}
\def \setQj{\left\{ Q_j \right\}}
\def \setQk{\left\{ Q_k \right\}}
\def \Ebar{\overline{E}}
\def \ebar{\overline{\epsilon}}
\def \eibar{\overline{\epsilon_i}}
\def \libar{\overline{\ell_i}}
\def \ebarN{\overline{\epsilon}_{\mbox{\tiny N}}}
\def \en{\epsilon_{\mbox{\tiny N}}}
\def \lbar{\overline{\ell}}
\def \Tg{T_{\mbox{\tiny G}}}
\def \Tf{T_{\mbox{\tiny F}}}
\def \Ta{T_{\mbox{\tiny A}}}
\def \Tc{T_{\theta}}
\def \Qc{Q_{\theta}}
\def \FH{F_{\mbox{\tiny HOMO}}}
\def \kH{k_{\mbox{\tiny HOMO}}}
\def \kB{k_{\mbox{\tiny B}}}
\def \wq{\omega_{\mbox{\tiny Q}}}
\def \Qast{Q^{\ast}}
\def \JQ{ {\cal J}^{\star}\left( Q\right)}
\def \cN{c_{\mbox{\tiny N}}}
\def \Ecore{E_{\mbox{\tiny C}}}
\def \Fcore{F_{\mbox{\tiny C}}}
\def \Fbarcore{\overline{F}_{\mbox{\tiny C}}}
\def \Shalo{S_{\mbox{\tiny H}}}
\def \pcore{p_{\mbox{\tiny C}}}
\def \Sroute{S_{\mbox{\tiny ROUTE}}}
\def \Qical{{\cal Q}_i}
\def \Tent{\mbox{Tent}}
\def \e{\epsilon}
\def \lK{\ell_{\mbox{\tiny K}}}
\def \l0{\ell_{\mbox{\tiny 0}}}
\def \bA{b_{\mbox{\tiny A}}}
\def \db{\delta b}

\newcounter{saveeqn}%
\newcommand{\alpheqn}{\setcounter{saveeqn}{\value{equation}}%
\stepcounter{saveeqn}\setcounter{equation}{0}%
\renewcommand{\theequation}
      {\mbox{\arabic{saveeqn}\alph{equation}}}}%
\newcommand{\reseteqn}{\setcounter{equation}{\value{saveeqn}}%
\renewcommand{\theequation}{\arabic{equation}}}%

\title{\Large\bf 
Speeding protein folding beyond the G\={o} model: How a little
frustration sometimes helps.
}
\author{{\large Steven S.~Plotkin }\\ 
Department of Physics, University of California, San Diego
}

\maketitle
\begin{center}
\bf Abstract
\end{center}
\par
\bigskip
\noindent
\begin{abstract}
Perturbing a G\={o} model towards a realistic protein Hamiltonian by
adding non-native interactions, we find that the folding rate is
in general  enhanced as ruggedness is initially increased, as long as
the protein is sufficiently large and flexible. Eventually the
rate drops rapidly towards zero when ruggedness significantly slows
conformational transitions.
Energy landscape arguments for thermodynamics and kinetics are coupled
with a treatment of non-native collapse to elucidate this effect.
\end{abstract}

\section{Introduction}

Theorists seek to capture the essence of protein folding with simple
models of a self-interacting polymer 
chain~\cite{GoN83,WolynesPG92,BryngelsonJD95,DillKA95:ps,VeitshansT97,Onuchic97,ShakhnovichEI98:rev,Dobson98,GarelT:rev98,Pande00RMP}.
There are two distinct limits pertaining to the nature of the interactions in
this minimalist approach. One is that of purely random interactions,
and is
considered too frustrated to describe real proteins.
Another is the G\={o} 
model~\cite{GoN75}, where the polymer is self-attractive only for those
parts of it in their native configurations. This is considered too
unfrustrated to describe real proteins, and also impossible to achieve in
practice. As these two models bracket the behavior of 
real proteins, we consider perturbing from the G\={o} model towards 
real protein interactions by adding some non-native
heterogeneity. Some of the effects of adding non-native interactions on
the folding mechanism for the Honeycutt-Thirumalai $\beta$-barrel
model~\cite{HoneycuttJD90} were 
investigated in~\cite{NymeyerH98:pnas,SheaJE98:jcp}.
At first glance one would expect that adding frustration begins to slow
the rate at the transition temperature,
or at best has initially no effect.
What follows is a derivation of the somewhat
counterintuitive result that in general the folding rate initially
increases as ruggedness is increased from zero.
Eventually of course the rate decreases drastically, so a 
plot of the folding rate {\it vs.} the amount of non-native heterogeneity
should look like fig.~(\ref{rate}). Then the question of where
real protein interactions reside on this plot may be addressed. 
For some fast-folding proteins, it is possible that 
non-native noise in  the system may actually assist folding.

\section{Thermodynamics}
Consider first the thermodynamics of a protein obtained from a
statistical analysis of a 
correlated landscape~\cite{PlotkinSS97}. The energy,
entropy, and free energy as functions of the fraction of native
contacts $Q$, are given by~\footnote{We will generally set
Boltzmann's constant $\kB = 1$  in this paper, so temperatures have
units of energy, and entropies are in units of $\kB$.}
\alpheqn
\begin{eqnarray}
E(Q) &=& Q \En - \frac{\Delta^2(Q)}{T}(1-Q)  \label{eq:E} \\
S(Q) &=& S_c (Q) - \frac{\Delta^2(Q)}{2 T^2}(1-Q) \label{eq:S} \\
F(Q) &=& Q \En - T S_c (Q) - \frac{\Delta^2(Q)}{2 T}(1-Q) \label{eq:F}
\end{eqnarray}
\reseteqn
These quantities are shown in figure~(\ref{s}), and the parameters
used in them are given in table I. 
$S_c(Q)$ in eq.~(\ref{eq:S})is the configurational entropy in the
system {\it vs.} $Q$, $\En$ is the extra internal energy in the native
state (the stability gap), and $\Delta^2(Q)(1-Q)$ is the non-native
variance which is a measure of the overall ruggedness of the energy
landscape (see eq.~\ref{eqV}).  The variance $\Delta^2(Q)(1-Q)
\rightarrow 0$ as $Q\rightarrow 1$. 
In obtaining the functional form of the non-native ruggedness, it is
assumed here that all the native contacts have roughly the same
strength.~\footnote{When there is variance in the native energies, the
non-native ruggedness terms are proportional 
to $(\Delta^2(Q)+ Q \Delta_{\mbox{\tiny N}}^2 (Q) )(1-Q)$, where
$\Delta_{\mbox{\tiny N}}^2$ is the native
variance~\cite{PlotkinSS98}. If the set of native energies has variance
but the distribution is fully specified, then the ruggedness terms are again
proportional to $\Delta^2(Q)(1-Q)$~\cite{PlotkinSS00:pnas}.}

The native energy is the number of native contacts $M$ times
the mean native attraction energy $\e$ ($\e < 0$). If $N$ is the number of
interacting residues in the polymer chain and $z$ is the number of
effective bonds per residue,
\be
\En = M \e = z N \e \: .
\label{eqEn}
\ee
The scale for the overall non-native ruggedness $\Delta^2(Q) (1-Q)$ is
given by 
\be
\Delta^2(Q) =  M b^2 \eta(Q) \: ,
\label{eqV}
\ee
where $\eta(Q)$ is the non-native packing density ($0< \eta(Q) <1$),
$b^2$ is the intrinsic variance per non-native interaction, and 
$M$ is the total possible number of (non-native) interactions,
i.e. the native state is assumed to be fully collapsed with the
maximal number of contacts, and this is the maximal number of
total contacts of any state.
The density $\eta(Q)$ tends to increase upon folding (see
section~\ref{sect:collapse}), hence the  
ruggedness scale $|\Delta|$ increases as well.
The strength $b$ of non-native interactions is taken 
to be weak:
\be
\frac{b}{\e} << 1 \: ,
\label{lesss}
\ee
therefore the ratio of folding transition temperature $\Tf$ to thermodynamic
glass temperature $\Tg$ is large
\be
\frac{\Tf}{\Tg} >> 1 \: ,
\label{tftg}
\ee
i.e. the proteins we consider are strongly (but not infinitely)
unfrustrated- we are perturbing away from the G\={o} model.

The configurational entropy $S_c(Q)$ has the property that 
entropy loss on folding is more rapid initially than in later stages.
We approximate this effect here by assuming the  form
\be
S_c(Q) = S_o (1-Q) - \Tent (Q)
\label{eqSo}
\ee
where $S_o \equiv N s_o$ is the total conformational entropy in the unfolded
($Q=0$) state ($s_o$ is the log number of conformational states per
residue), and  
$\Tent(Q)$ is a tent function:
\be
\Tent(Q) = \left\{ \begin{array}{ll}
		2 \phi Q & Q<1/2 \\
		2 \phi (1-Q) & Q > 1/2
	\end{array}
\right. \: .
\label{eq:tent}
\ee
We've let the barrier be at $\Qddag = 1/2$ for simplicity of argument.


At the transition temperature $\Tf$, the free energy of the folded and
unfolded states are equal:
\bea
F(0) &\cong& F(1) \nonumber \\
-\Tf S_o - \frac{\Delta^2(0)}{2 \Tf} &\cong& \En \: .
\label{eqTf}
\eea
Using this relation in eq.~(\ref{eq:F}) gives
\be
\left. \frac{F(Q)-F(0)}{T}\right|_{\Tf} = \Tent(Q) 
- \frac{Mb^2 (1-Q)}{2 \Tf^2} \left[
\eta(Q)-\eta(0) \right] \: .
\label{fattf}
\ee
When $b=0$ the free energy at $\Tf$ is the tent function (see
fig.~\ref{s}), and so $\phi$ in eq.~(\ref{eq:tent}) is thus $\Fdag
(b=0)/\Tf$. Then from eq.~(\ref{fattf}) the free energy barrier at
$\Tf$ is given by
\be
\frac{\Delta F^{\neq}}{\Tf} \approx \frac{\Delta F^{\neq} (b=0)}{\Tf}
- \frac{M b^2}{4 \Tf^2} \Delta \eta^{\neq} \: ,
\label{eq:bar}
\ee
where $\Delta \eta^{\neq} = \eta(\Qdag=1/2)-\eta(0)$ is the change in
non-native density between the barrier peak and unfolded state
($\Delta \eta^{\neq} < 1$).
So long as $\Delta \eta^{\neq} > 0$, the barrier height
decreases as non-native heterogeneity ($b^2$) increases, as shown in
fig.~(\ref{s}).
We now show that this is nearly always the case, by considering the physics
of collapse for our problem in question.

\subsection{The Collapse Transition}
\label{sect:collapse}

In this section we investigate the coupling of non-native density with
the amount of native structure present in a protein, by showing that
native topological constraints can induce 
a collapse transition on the non-native parts of the protein.
Then the trend in eq.~(\ref{eq:bar}) of adding non-native ruggedness would
be to lower the folding barrier.

Collapse occurs below a temperature $\Tc$, defined as the temperature
where the free energy of the coil and collapsed molten globule phases 
(both at $Q\cong 0$) are equal:
\be
F_{coil} (\Tc) = F_{mg}  (\Tc) \: .
\ee
Again using eq.~({\ref{eq:F}), but now noting that the
conformational entropies are 
different in the coil and globule phases, and that
$\eta \cong 1$ in the globule and $\eta \cong 0$ in the coil phase,  we have 
\be
- \Tc S_{coil} = - \Tc S_{mg} - \frac{Mb^2}{2\Tc} - M a \: .
\ee
Note we have now allowed for a mean homopolymer attraction $a$ in
general, for reasons which will become clear below.
Using the reduction in entropy for collapsed {\it vs.} coil chain
statistics~\cite{BryngelsonJD90,Nemirovsky92}
\be
S_{coil} -  S_{mg} = N \log \nu - N \log (\nu/e) = N
\ee
gives for the collapse temperature
\be
\Tc = \frac{z a}{2} \left( 1 + \sqrt{1+\frac{2 b^2}{z a^2}} \right) \:
.
\label{tc1}
\ee
Note from~(\ref{tc1}) that when $b=0$, $\Tc = za$, i.e. the collapse
temperature is the mean homopolymer attraction times the number of
contacts per residue, and when $a=0$, $\Tc = b\, \sqrt{z/2}$,
i.e. non-native heterogeneity can drive collapse, with the 
collapse temperature now scaling with the root number of contacts per
residue times the width of interactions.

Now we note that in our model (G\={o} perturbed by weak heterogeneity,
with $a=0$)
collapse and folding  will tend to occur together, with folding
driving the collapse 
through native structural constraints. So the total density 
increases from zero to one as $Q \rightarrow 1$, and the non-native
density $\eta(Q)$ should increase as well since non-native polymer is
more strongly constrained by larger native cores, see
figure~\ref{fig:corehalo}.
The simplest approximation to capture this
increase in density upon folding is to replace the 
mean homopolymer field $a$ by the native energy scale $\e$ times the
fraction of native bonds made $Q$:
\be
a(Q) = \e Q \: .
\label{eq:a}
\ee
This is the effective homopolymer
field for the ensemble of states with fraction $Q$ of native structure.
Using~(\ref{eq:a}) in~(\ref{tc1}) and noting that the glass
temperature 
\be
\Tg = \sqrt{\frac{z b^2}{2 s_{mg}}}
\ee
gives
\be
\Tc = \Tf \left( \frac{s_o Q}{2} + \sqrt{ \left( \frac{s_o
Q}{2}\right)^2 + s_{mg} \left(\frac{\Tg}{\Tf}\right)^2 } \right)
\label{tc2}
\ee
where $s_o$ and $s_{mg}$ are the entropy per residue in the coil and
globule state respectively.
Note that in eq.~(\ref{tc2}) $\Tc > \Tf$ as long as the term in
parentheses is greater than one. This gives a critical value $\Qc$
where collapse occurs during folding, i.e. when $Q\gtrsim \Qc$, $\eta
\approx 1$ 
and when $Q \lesssim \Qc$, $\eta \approx 0$. This is sketched in
fig~(\ref{eta}A) below. Solving for $\Qc$ gives
\bea
\Qc &=& \frac{1}{s_o} - \frac{s_{mg}}{s_o}
\left(\frac{\Tg}{\Tf}\right)^2 = \frac{1}{s_o} - \frac{s_o b^2}{2 z
\e^2} \cong \frac{1}{s_o} 
\label{eq:qc} \\
\Tc(Q) &\cong& Q s_o \Tf = Q z \epsilon
\label{eq:qc2}
\eea
since, by construction of the problem, eqn.~(\ref{lesss}) holds,
e.g. say 
$b/\e$ is about $1/20$. Then the second term in~(\ref{eq:qc}) is of
order $1/400$ and can be neglected. Equation~(\ref{eq:qc}) says that
the more chain entropy the polymer has (the more flexible it is) the
{\em sooner} it collapses when folding at the transition temperature.

Calorimetric measurements of the conformational entropy change per residue
in unfolding to the coil state for say barnase give $s_o \cong 55 \:
\mbox{J}/\mbox{K} \cdot \mbox{mol residue} \cong 6.8 \kB$ per
residue~\cite{MakhatadzeGI96}.  
This entropy also counts side chain conformational
entropy, which is estimated to be about $13 \: \mbox{J}/\mbox{K} \cdot
\mbox{mol residue} \cong 1.6 \, \kB$ per residue~\cite{DoigAJ95}, giving
a net chain conformational entropy of about $5.2 \, \kB$ per residue in
the coil state, and therefore  
$\Qc \cong 0.2$. 
For typical off-lattice simulations~\cite{NymeyerH98:pnas,ClementiC00:jmb}
$s_o \cong 3.4 \kB$, therefore $\Qc \cong 0.3$. So collapse typically 
occurs before the barrier is reached (see fig.~(\ref{eta})C).
For lattice
simulations $\Qc \cong 0.6$, which is around $\Qdag$. 
In any event,  $\eta(\Qddag)$ will tend to be greater than $\eta(0)$
as long as the system is large enough and bulk thermodynamics can be
used (however see caveats in appendix~\ref{appA}), see figure~\ref{eta}. 
The values of non-native packing density obtained from simulations
appear to be smaller than one, probably because of finite size and
stiffness effects. Applying bulk thermodynamics to the residual segments of
non-native polymer may not be an accurate approximation in some cases.

More complete treatments of the coupling of density with native
similarity can be made within the energy landscape
framework~\cite{PlotkinSS97}. It is fairly 
straightforward to write an approximate free 
energy as a function of both $\eta$ and $Q$ and then minimize with
respect to $\eta$ to obtain the density as a function of $Q$. We have
taken the simplest approach here to illustrate the coupling of
collapse with the thermodynamics.
Some cautionary notes are
made in Appendix A regarding a possible reversal of the trend on
barrier height in
small, stiff proteins, or proteins with a significant amount of
generic self-attraction.

\section{Kinetics}

What about the rate? The question is now
whether the increase in prefactor is larger  than the decrease in
barrier height, as we add non-native heterogeneity. Since the
ruggedness is weak, the kinetics are single exponential (there is a
single dominant folding barrier), and a Kramers law holds for the rate: 
\be
k \cong \tau^{-1}(b)\; \mbox{e}^{-\Delta \Fdag (b)/T} \: ,
\label{rate1}
\ee
where the prefactor is proportional to the reciprocal of the
reconfiguration time
scale~\cite{BryngelsonJD89,SocciND96:jcp,WangPlot97,PlotkinSS98}.  

If we were to follow the  argument for the dependency of the
prefactor on ruggedness for an uncorrelated
landscape~\cite{BryngelsonJD89}, or for a correlated landscape at
low temperature with 
activated dynamics~\cite{WangPlot97}, we would find that the
ratio of rates 
\be
\left. \frac{k(b)}{k(b=0)}\right|_{\Tf} = \exp\left( \frac{M b^2}{4
\Tf^2} \Delta\eta^{\neq} - \frac{M b^2}{\Tf^2} {\cal G} \right)
\label{eq:k1}
\ee
where ${\cal G}$ is a function of  $T/\Tg$ on the uncorrelated
landscape, and on the correlated landscape
is a function of both $T/\Tg$ and structural entropic factors having
to do with the density of states of given similarity to a trap.
So by inspection of eq.~(\ref{eq:k1}), in this low temperature regime
the rate may go up or may go down 
with non-native interaction strength.

However an important result arising from energetic correlations in the
landscape 
is the existence of a critical temperature $\Ta$ where the dynamics
becomes unactivated~\cite{WangPlot97}. Above this temperature the dynamics is
similar to reconfigurations in a normal liquid rather than the hopping
dynamics of trap escape in a supercooled liquid or glassy
system~\cite{Kirkpatrick87b,Gotze91,WangPlot97,TakadaS97:pnas}, i.e. 
above $\Ta$, the prefactor $\tau^{-1}(b)$ remains nearly constant with
ruggedness, since at these temperatures the Rouse modes depend much
more weakly on the ruggedness introduced.
The existence of such a temperature scale can be seen from the
following simple argument~\cite{WangPlot97}. We can think of escape
from a trap as a 
mini-unfolding event: escape is driven by entropy and is opposed by
the putative trap's low energy, say $E_i$. Then, as in unfolding, the escape
barrier arises from a mismatch between entropy gains and energy
losses as the system reconfigures out of the trap, so we can rewrite
equation~(\ref{eq:F}) for the free energy 
relative to the state $i$ as
\be
F(q ) = q E_i - T S_c (q) - \frac{\Delta^2}{2T} (1-q)
\label{eq:trap}
\ee
where $\En$ in~(\ref{eq:F}) is replaced by $E_i$, $Q$
in~(\ref{eq:F}) is 
replaced by $q$, defined as the fraction of contacts shared with state
$i$, and 
density changes during untrapping are not particularly 
important since $\Ta < \Tc$ (energetic trapping occurs only when  the
polymer is collapsed~\cite{BryngelsonJD90}). 
The transition to unactivated dynamics occurs when the
states typically occupied at that temperature have zero escape barrier. 
Setting $E_i$ in~(\ref{eq:trap}) to the thermal energy
of states at temperature $T$,~\footnote{c.f. eq.~(\ref{eq:E}) at
$Q=0$. For strata of states with  $Q > 0$ the larger
ruggedness scale $\Delta (Q)$  increases $\Ta$ and $\Tg$ for
that stratum of states.} 
\be
E_i \approx \overline{E}(T) = -\frac{\Delta^2}{T}
\label{etherm}
\ee
we note that $\Ta$ occurs when the free 
energy profile is downhill away from the trap at $q=1$, i.e when $\D
F_{>} / \D Q =0$ in our model, where the subscript $>$ indicates the
high $q$ region of the piecewise free energy
function~(\ref{eq:trap}) (eq.~(\ref{eqSo}) for the
configurational entropy has a piecewise structure). Using
equations~~(\ref{eq:tent}),~(\ref{eq:trap}), and~(\ref{etherm}), this gives
\be
\Ta  = \Tg \left(1 - \frac{2 \phi}{S_o}\right)^{-1/2} 
\label{ta}
\ee
for the transition temperature to  activated dynamics. From~(\ref{ta}) we
see that $\Ta > \Tg$ by an amount which depends on the deviation from
linear entropy loss over the total unconstrained entropy, i.e. by the
entropic contribution to the barrier. There is no energetic
contribution since we have used $q$ as the order parameter and assumed
pairwise interactions.~\footnote{
One must be consistent in interpreting eq.~(\ref{ta}). In mean-field
theory, $\phi$ is extensive and $\Ta > \Tg$  in the thermodynamic
limit. But in the capillarity theory the entropic deviation $\phi$
comes from surface entropy and should scale as $N^{2/3}$ (or even with
a smaller power if the interface is roughened). One might argue that
since $S_o \sim 
N$, $\Ta$ approaches  $\Tg$ as $N \rightarrow \infty$, but
matching the theories in this fashion is incorrect since
eq.~(\ref{eq:trap}) is not valid in the capillarity limit.
In the capillarity theory a dynamical transition can only be seen by
investigating where the intensive surface tension vanishes.}

For typical size proteins of $N\sim
100$, $\Ta \approx (1.6\,-\, 1.8)\, \Tg$.
A plot of the escape time on a correlated landscape is
given below (see fig.~\ref{tau}). 
In the regime we are interested in, $\Tg / \Tf <<1$
(c.f. eq.~(\ref{tftg})), so it is also true that
\be
\frac{\Ta}{\Tf} << 1
\ee
and so the characteristic temperatures where folding occurs ($\sim
\Tf$) are way above the transition temperature for activated diffusion by
construction of the problem; see figure~(\ref{tau}).

Expanding eq.~(\ref{rate1}) around $b= 0$ using eq.~(\ref{eq:bar})
gives the rate for weak non-native heterogeneity:
\be
\frac{k}{k_{\mbox{\tiny GO}}} \cong 1 + M \, \Delta \eta^{\neq} \,
\frac{s_o}{4 z^2}  \, \left( \frac{b^2}{\e^2}\right)  \: ,
\ee
where $\Delta \eta^{\neq} = \eta(\Qdag) - \eta(0)$ and we've used
$\Qdag \cong 1/2$. 

The increase in rate occurs until around $\bA - \db(N)$, where $\bA$
is where $\Tf(b) \cong \Ta(b)$ (about $0.3$ here), and $\db(N)$ is the
finite-size fluctuation of $\bA$ due to temperature fluctuations
which round the transition~\cite{Landau80}. That is,
\be
\db \approx \delta \Ta = \left. \frac{T}{C_v}\right|_{\Ta}
= \left. \frac{T^2}{\sqrt{M b^2}}\right|_{\Ta} \approx \frac{\bA}{\sqrt{N}} 
\ee
where eq.~(\ref{eq:S}) was used for the entropy at $Q=0$. This gives
a value for $\db \approx 0.02 - 0.04$.

Realistic values of $b^2/\e^2$ for a typical protein can be obtained
from the ratio of 
folding to glass temperature, given by 
\be
\Tf/\Tg = \lambda + \sqrt{\lambda^2 -1}
\ee
where $\lambda = \sqrt{z/2 s_o} (\e/b)$~\cite{GoldsteinRA-AMH-92}.
Commonly accepted values of $\Tf/\Tg$ for proteins are about $1.6 -
2.0$. Using the values in table I for $s_o$ and $z$ gives
\be
\left.\frac{b^2}{\e^2}\right|_{\mbox{proteins}} \approx (0.1 \, - \,
0.15)
\ee
which is above the rate enhancement regime (see fig.~\ref{rate}). If 
the effects of non-native rate enhancement are observable, they
will be seen possibly in only the fastest folding proteins.
On the other hand such an observation would support the existence of a
dynamic glass transition in protein systems.

\section{Conclusion}
As non-native heterogeneity is 
increased from zero, the folding rate initially increases by a factor
of about $2- 4$, then eventually drops rapidly towards zero
 (see fig.~\ref{rate}). There is a regime near the G\={o} Hamiltonian
where the rate 
is relatively robust to changes in non-native interaction strength.
The density of non-native polymer must be greater at the barrier peak
than in the unfolded state for rate enhancement to be observed.
Why does the rate initially increase?
The upshot is as follows.
First, it follows from energy landscape theory that if there is
no change in density upon folding, then changes in ruggedness do not
affect the barrier height for a well-designed protein
(c.f. eq.~(\ref{eq:F})), that is, effects on the barrier must come
from the coupling of non-native density to the degree of nativeness.
Moreover, the strength of the
ruggedness per residue increases with non-native density since there are simply
more interactions, and non-native density tends to
increase with nativeness, at least for weak non-native interaction
strength. Then, since ruggedness lowers the free energy, the free
energy at the barrier position is lowered more than the unfolded free
energy. So as non-native interaction strength is increased from zero,
the barrier lowers and the rate increases if the effect on the
prefactor is weaker. But the prefactor is related to the
reconfigurational diffusion
time~\cite{BryngelsonJD95,SocciND96:jcp,WangPlot97}, and since a
dynamic glass transition 
is expected in such systems, there will be a window for weak
ruggedness within which the diffusion time is relatively unaffected as
ruggedness is increased. Thus the rate initially increases, as shown
in fig.~\ref{rate}.
This phenomena provides a good example of how energy landscape theory
can be applied to the physics of protein folding to reveal and explain
a counter-intuitive result.

\subsection{Acknowledgments}
We thank Jose' Onuchic and Cecilia Clementi for many
enjoyable and insightful 
discussions, and Peter Wolynes specifically for helpful discussions on
the collapse transition. We are also grateful to Hugh Nymeyer for
providing the off-lattice simulation data and constructive advice. 
This work was supported by National
Science Foundation (NSF) Bio-Informatics Fellowship DBI9974199 and
by NSF Grant 0084797.

\section{Appendix A: Caveats due to finite-size, generic attraction,
and stiffness effects} 
\label{appA}

The derivation leading to eq.~(\ref{eq:bar}) assumed mean-field theory
could be applied, that the protein could could be treated as a
bulk system, and that properties arising from chain connectivity would
not alter the results arising from the energetics in the problem.
In particular we have assumed that the polymer persistence length or
Kuhn length $\lK$ is much less than the length of a typical piece of
disordered protein $\l0$ near the barrier peak, so that a dangling
piece of disordered polymer may interact with itself.
If the protein under study is particularly stiff and/or small, the
return length of polymer fragments may be comparable to the length of
the disordered pieces, reducing the number of non-native interactions
near the barrier peak relative to the number in the unfolded state.
Then in eq.~(\ref{fattf}}) the density $\eta(\Qdag) \lesssim \eta(0)$
and no reduction in barrier height with non-native interactions would be
seen. 

For the $27$-mer lattice model $\lK \approx 3-4$; these models are
relatively stiff compared to their total length. For typical off-lattice
models on the other hand, $\lK \approx 3$ but they are considerably
longer, e.g. for SH3, $N=57$.
If all the non-native polymer is in one strand, $\l0
\approx N/2$ at the barrier peak. Then $\lK/\l0 \approx 0.26$ for the
lattice model, and $\lK/\l0 \approx 0.11$ for the SH3 off-lattice model.
If the non-native polymer is distributed among a number of
disordered strands that can dress the native core, 
roughly $(1/6)\times N^{2/3}$~\cite{FinkelsteinAV97}, then $\lK/\l0
\approx 0.40$ for the lattice model, and $\lK/\l0 \approx 0.26$ for
the off-lattice model.

Neither of these numbers are very small indicating that the collapse
transitions are quite rounded, and the effect on folding rate will be
mild if it exists. 
In fact the discrepancy of $\lK/\l0$ between the off-lattice and
on-lattice models, although fairly small here, leads to different
behavior of the non-native density $\eta(Q)$, as shown in
fig.~\ref{eta}.  In the lattice system
$\eta(\Qdag) \lesssim \eta(\Qu)$, but in the off-lattice system
$\eta(\Qdag) \gtrsim \eta(\Qu)$. Hence we anticipate the
rate-enhancement effects will be seen in off-lattice models, but
probably not in at least the  shorter on-lattice
models~\cite{ClementiC00:un}.

Real proteins may tend to have some net homopolymer attraction
inducing generic collapse. This decreases the change in density upon
folding and would further attenuate any rate enhancement effect 
present. However at least some proteins are sufficiently stable that 
collapse and folding are concomitant~\cite{PlaxcoKW99:nsb}.
Moreover G\={o} models, for which collapse and folding are concomitant
by construction, give reasonably accurate predictions of $\phi$-values
and barrier
heights~\cite{ShoemakerBA97,ShoemakerWang99,FinkelsteinAV99:pnas,AlmE99,MunozV99,ClementiC00:jmb,ClementiC00:pnas}.
Collapse accompanies folding when the folding transition temperature
$\Tf$ given by 
eq.~(\ref{eqTf}) is comparable to the collapse temperature $\Tc$ in
eq.~(\ref{tc1}). For weak ruggedness,
\be
\frac{\Tf}{\Tc} \approx \frac{\e}{a s_o} \left( 1 - \frac{s_o}{2 z}
\frac{b^2}{\e^2} - \frac{1}{2 z} \frac{b^2}{a^2} \right) \, .
\ee
So the effects of generic collapse are not important in the problem
as long as $\e \gtrsim a \, s_o$, to the first approximation.

Several additional features may affect the folding rate.
In finite-sized systems the unfolded state tends to have partial
order. Moreover its position may drift, along with that of the
transition state, as non-native variance is increased. This modifies
the barrier height.
Additionally the density $\eta(Q)$ in figure~\ref{eta} is exact in the limit
$b\rightarrow 0$, but for nonzero $b$, $\eta(Q)$ may begin to alter in
structure. Accounting for these effects will modify the folding rate,
but shouldn't alter the general trend of figure~\ref{rate}.

\newpage
\onecolumn
\begin{table}[t]
\begin{minipage}{1.0\linewidth}
\caption{\scriptsize
PARAMETERS IN THE MODEL
}
\centering
\begin{tabular}{cccccccc}
{} & {Polymer } & {Contacts  } & {Conformational} & {Entropy} 
& {Native} & {R.M.S.} & {Folding}\\
{} & {length} & {per } & {entropy} & {nonlinearity} & {contact} 
& {non-native} & {transition} \\
{} & {} & {residue} & {per residue} & {(eq.~\ref{eq:tent})} & {energy} 
& {contact} & {temperature}\\
{} & {} & {(eq.~\ref{eqEn})} & {(eq.~\ref{eqSo})} & {} & {(eq.~\ref{eqEn})} 
& {energy} & {(eq.~\ref{eqTf})}\\
{} & {} & {} & {} & {} & {} & {(eq.~\ref{eqV})} & {} \\
\phantom{re}(Model) & \phantom{re}($N$) & \phantom{re}($z$) & 
\phantom{re}($s_o$)
& \phantom{re}($\phi$) & \phantom{re}($\epsilon$) & \phantom{re}($b$) 
& \phantom{re}($\Tf$) \\
\hline
 G\={o}, G\={o} + Non-native    & 64  &	1.25 & 3.4 & 5.0 & -1.0 & 0 , b & 0.37  

\end{tabular}
\label{table1}
\end{minipage}
\end{table}

\subsection{Figure Captions}

\begin{figure}[htb]
\hspace{.7cm}
\psfig{file=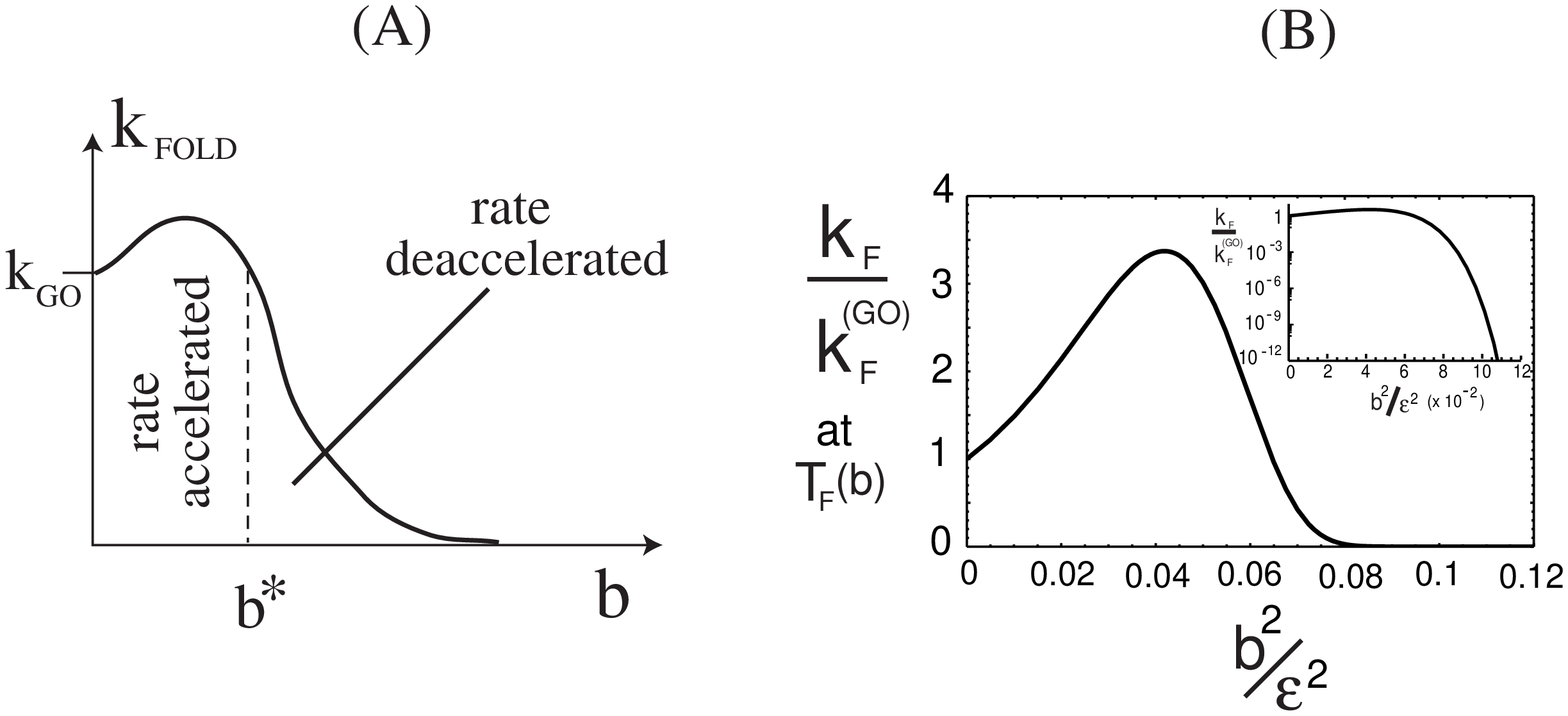,height=14cm,width=6cm,angle=-90}
\caption{}
\label{rate}
\end{figure}

\begin{figure}[htb]
\hspace{.7cm}
\psfig{file=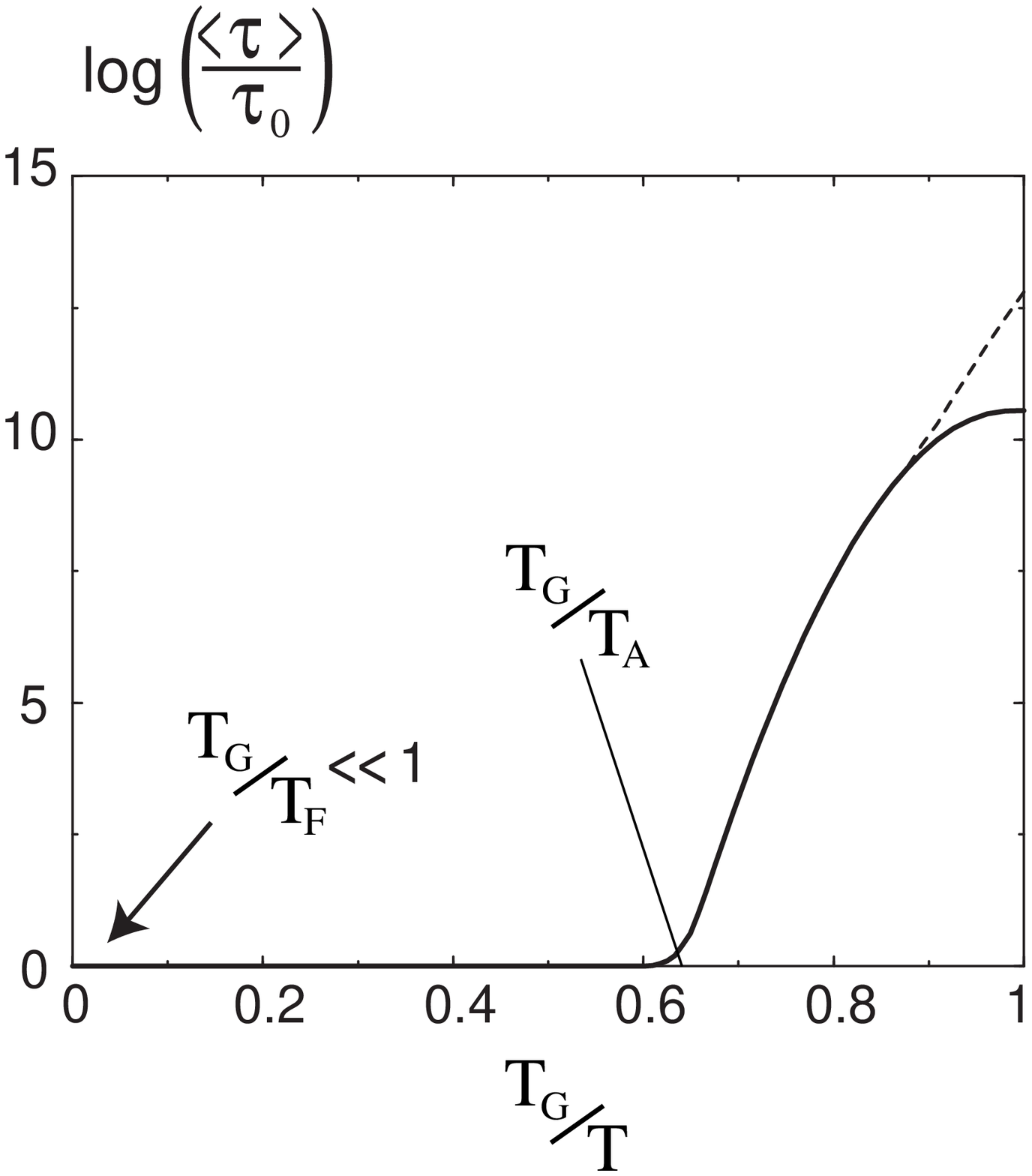,height=6cm,width=6cm,angle=0}
\caption{}
\label{tau}
\end{figure}

FIG.~\ref{rate} $\;\;$
Rate {\it vs.} non-native heterogeneity is split up into two
regimes, one where it assists folding, the other where it
hinders. The rate plotted here is the folding rate at $\Tf$ which is
itself a function of $b$ (c.f. eq.~(\ref{eqTf}), however $\Tf$ changes
by only $\sim 1 \%$ over the range of this plot.
(a) Schematic depicting the two regimes, (b) Result of the theoretical
model introduced in the text (see equations~\ref{eq:bar},~\ref{rate1},
and figure~\ref{tau}) for a system with parameters given in
table~I. Initially the rate rises as $\sim N b^2/\e^2$, then strongly
decreases for larger $b$ as non-native interactions slow
conformational transitions. 
Realistic protein interactions are believed to have typical values of
$b^2/\e^2 \approx 0.1 - 0.15$~\cite{OnuchicJN95:pnas}, which is above
the rate enhancement regime. The inset of (B) shows a semi-log plot of
the same rate; it can be seen that there is a regime where the rate is
roughly constant as ruggedness is increased from zero, then a turnover
where the rate drastically decreases.

\begin{figure}[htb]
\hspace{.7cm}
\psfig{file=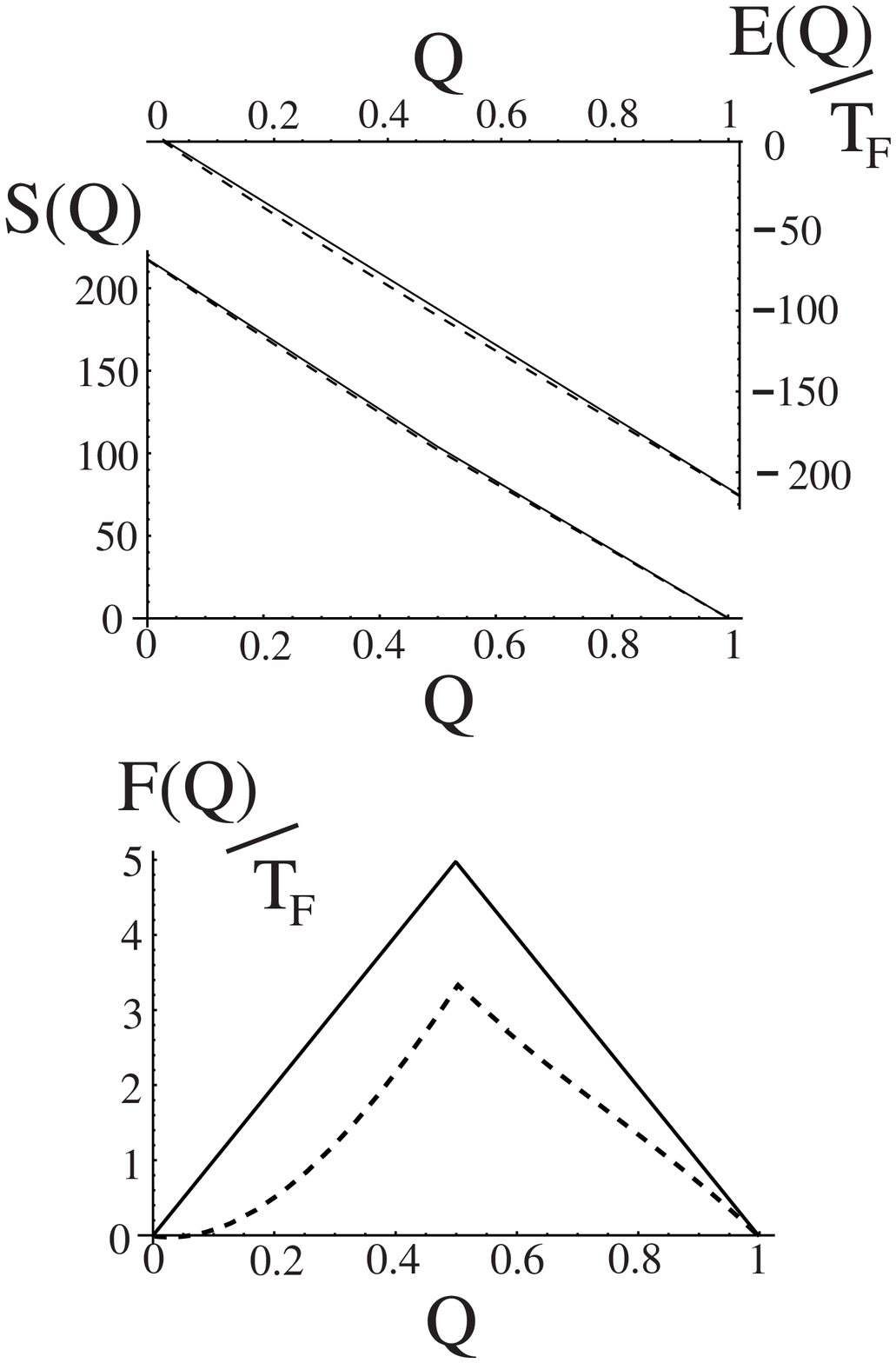,height=10cm,width=6cm,angle=0}
\caption{}
\label{s}
\end{figure}

\begin{figure}[htb]
\hspace{.7cm}
\psfig{file=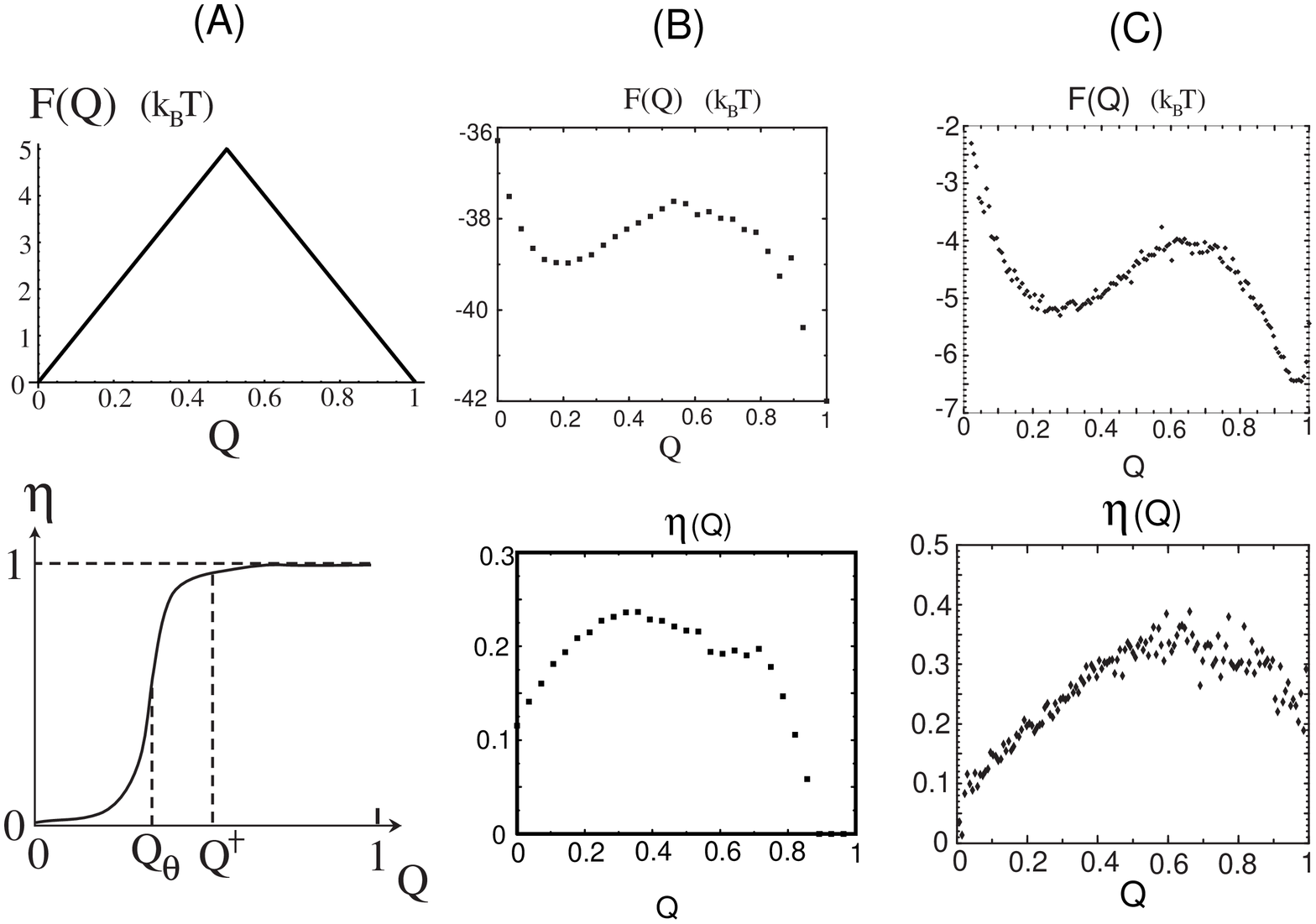,height=14cm,width=8cm,angle=-90}
\caption{}
\label{eta}
\end{figure}

FIG.~\ref{s} $\;\;$
Energy, entropy, and free energy {\it vs.} $Q$ in the model,
for the G\={o} model with $\Delta^2 =0$ (solid line), and when ruggedness is
introduced, when $\Delta^2 > 0$ (dashed line). We took $b^2 \cong
0.04$ (see eq.~\ref{eqV}), where the folding rate is maximal (see
fig.~\ref{rate}). Parameters used in the  
model are given in table~\ref{table1}.
A bilinear approximation for the configurational entropy is used
here, giving a tent functional form for the G\={o} free energy at the
transition temperature $\Tf$. The non-native density function
$\eta(Q)$ used here is a fit to the off-lattice data in fig.~\ref{eta}C.
The folding free energy barrier at $\Tf$ 
is {\it lowered} by non-native heterogeneity, because the energy is
lowered twice as much as the entropy (see eq.~1).

\begin{figure}[htb]
\hspace{.7cm}
\psfig{file=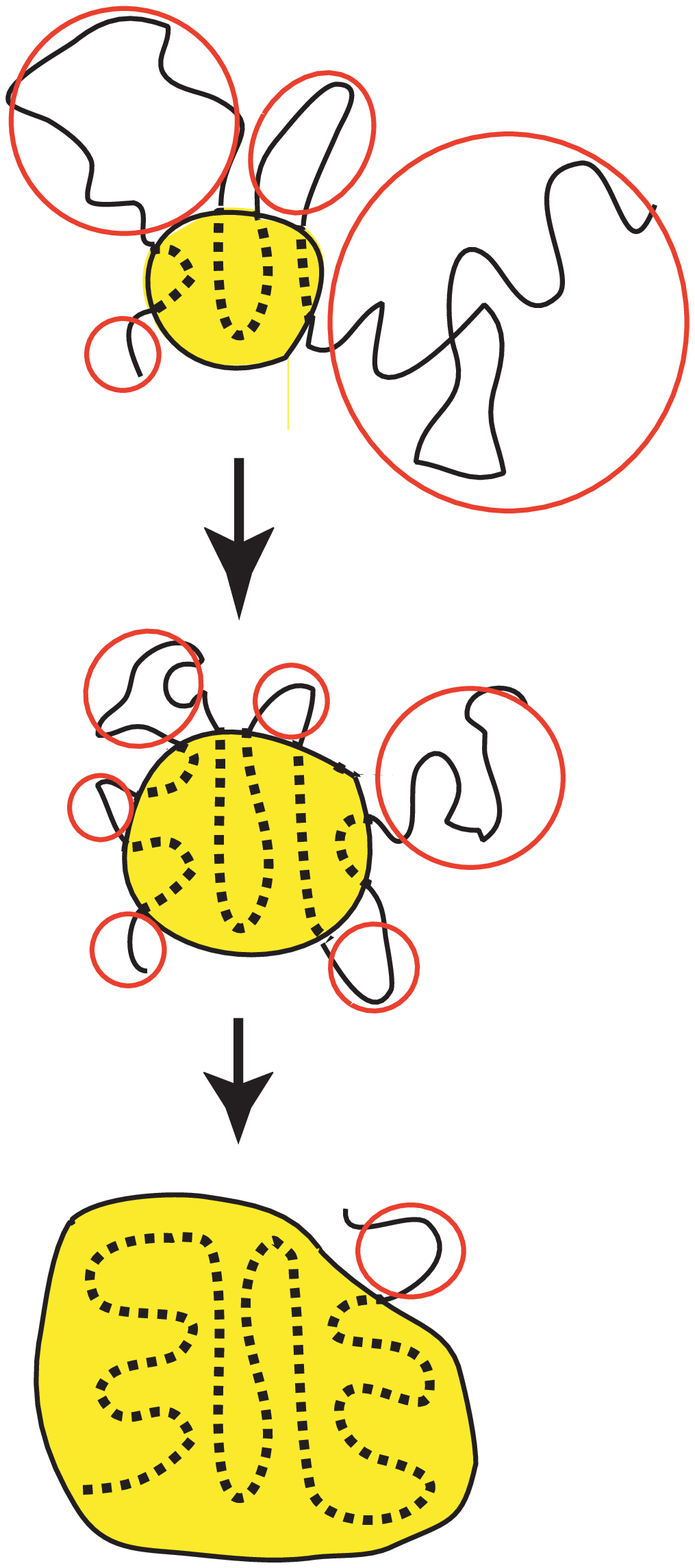,height=14cm,width=6cm,angle=0}
\caption{}
\label{fig:corehalo}
\end{figure}

FIG.~\ref{fig:corehalo} $\;\;$
As folding progresses, the non-native polymer halo
surrounding the native core (central shaded globule) 
has more topological constraints placed upon it. Therefore the
non-native packing density, given by the total number of non-native contacts
divided by the characteristic volume of
non-native polymer (open spheres), tends to increase. 

FIG.~\ref{eta} $\;\;$
Free energy profiles $F(Q)$ and non-native polymer density
$\eta(Q)$ in the model for (a) the simple bulk-mean-field model used
in the derivation, (b) a 27-mer lattice model, (c) an
off-lattice 
model for the $57$ residue fragment corresponding to the
$\alpha$-spectrin {\it SH3} domain (PDB code 1BK2). 
 (a) Collapse 
occurs at $\Qc$ before the barrier peak at $\Qdag$. The transition is
rounded for typical sized proteins, as in (b) and 
(c). In (b), the non-native density is overall fairly small, and is
comparable in the unfolded and transition state, $\eta(\Qu) \approx
\eta(\Qdag) \approx 0.22$. Thus eq.~(\ref{fattf}) gives a barrier
height roughly independent of $b$ at least for small $b$. In (c) on
the other hand, the non-native density rises to values larger in
overall magnitude, and is monotonically increasing until the barrier
peak: $\eta(\Qdag) \approx 0.35$ and $\eta(\Qu) \approx 0.2$. For the
parameter values in table I, eq.~(\ref{fattf}) then gives a barrier
height decreasing with $b$ as $\Delta \Fdag/\Tf \approx -10 \,
b^2/\e^2$. The drop-off 
at high $Q$ in (B) and (C) is most probably due to stiffness effects
on the small pieces of residual non-native polymer in this regime. 

FIG.~\ref{tau} $\;\;$
Log of the reconfiguration  time {\it vs.} reciprocal
temperature in units of 
$\Tg$, for a system of size $N=64$, adapted from
reference~\cite{WangPlot97}. This is used in equation~(\ref{rate1}) to
produce the rate curve in figure~\ref{rate}.
At temperatures above $\Ta$ the time to reconfigure is $\sim
\tau_o$, below $\Ta$ the time increases exponentially as  
$\exp( f(T/\Ta) N)$ with $f(x) = 0$ for $x<1$. 
The width of the transition $\rightarrow 0$ as
$N\rightarrow \infty$ and the value of $\Ta \rightarrow \approx 1.8$
as $N\rightarrow \infty$ for the mean field correlated landscape. 
When $T \lesssim
\Tg$ for a finite system, 
the deepest trap tends to dominate the kinetics, and the relaxation
rate turns over to an Arrhenius law with slope corresponding to the barrier
height for escape from that trap (shown schematically by the dashed line).

\newpage
\onecolumn

\newpage

\twocolumn

\newpage

\newpage

\onecolumn

\newpage

\twocolumn

\newpage
\pagestyle{plain}


\end{document}